\documentclass{article}
\usepackage[normalem]{ulem}
\usepackage[utf8]{inputenc}
\usepackage[T1]{fontenc}
\usepackage{lipsum} % for generating filler text
\usepackage[margin=1.5in]{geometry} % sets all margins to 1 inch
\usepackage{amsmath,amssymb,amsthm}
\usepackage{graphicx}
\usepackage{authblk}
\usepackage{xcolor}
\usepackage{placeins}
\usepackage{url}
\usepackage{caption}

\title{Machine-Learned Dynamical Representations for Accelerated RiteWeight Convergence}
\author[1]{Sagar Kania\thanks{Corresponding author: \texttt{sagarkania@gmail.com}}}

\affil[1]{Independent Researcher, Portland, OR 97214, USA}

\date{\today}

\date{\today}

\begin{document}

\maketitle
\section*{Abstract}

The increasing use of generative models has made ensembles of short molecular
dynamics trajectories increasingly common, creating a growing need for methods
that can recover physically meaningful steady-state populations and kinetics
from improperly weighted conformational ensembles. Randomized Iterative
Trajectory Reweighting (RiteWeight) addresses this problem through repeated
random clustering and iterative reweighting, without requiring the fixed
Markovian discretization used in conventional Markov state models (MSM).
However, the choice of reduced feature space in which RiteWeight performs
random clustering has not been systematically investigated. Here, we compare
two machine-learned representations, DeepTICA and SPIB-VAE, with linear TICA
for recovering steady-state observables from flawed distributions. DeepTICA learns nonlinear coordinates by targeting slow
transfer-operator eigenmodes, whereas SPIB-VAE compresses configurations into
a low-dimensional latent space while retaining information predictive of
future metastable states. DeepTICA provided a comparatively robust RiteWeight
representation under limited hyperparameter exploration, whereas SPIB-VAE
benefited more strongly from broader optimization. Moreover, a kinetic score
computed from a coarse MSM at a resolution comparable to that used for
RiteWeight random clustering provided a useful criterion for efficiently
selecting reduced representations and their hyperparameters for RiteWeight.

\section{Introduction}

Recent advances in protein structure prediction and generative modeling can generate diverse, physically plausible conformations spanning multiple regions of biomolecular configuration space. These structures can provide starting points for ensembles of short molecular dynamics trajectories, enabling broader exploration than may be accessible from a single conventional trajectory \cite{Bhakat2026BioEmu, Vani2023AlphaFold2RAVE, Tang2026GenCOMPAS}. However, neither the generated conformations nor the resulting trajectory ensemble is necessarily distributed according to the equilibrium or nonequilibrium steady-state distribution of interest. Additional statistical treatment is therefore required to recover correct configurational populations and associated thermodynamic and kinetic observables \cite{Bhakat2026BioEmu, Tang2026GenCOMPAS, Zhu2026TyrosineKinases}.

Conventional Markov state models (MSM) can analyze such trajectory ensembles, but their accuracy depends on a fixed discretization of configuration space and on choosing a lag time for which the dynamics between discrete states are sufficiently Markovian \cite{caflisch2011equilibrium,vitalis2019msm_bias, voelz2020adaptive,suarez2021markov}. In our previous work, we introduced Randomized Iterative Trajectory Reweighting (RiteWeight), which repeatedly changes the cluster boundaries and uses the resulting transition matrices to estimate stationarity rather than to propagate dynamics \cite{Kania2026RiteWeight}. RiteWeight thereby mitigates fixed-discretization error and does not require cluster-level Markovianity, enabling accurate reweighting using substantially shorter trajectory segments than are typically required for MSM analysis \cite{Kania2026RiteWeight}.

In the ideal limit of dense sampling and complete iterative convergence, the RiteWeight stationary solution is determined by the underlying fine-resolution transition process rather than by a particular coarse clustering. In practical calculations, however, both the available data and the number of RiteWeight iterations are finite. Configurations that remain close in the feature space used for random clustering are unlikely to be separated and may therefore remain insufficiently reweighted. Consequently, an effective RiteWeight representation should distinguish configurations separated by important kinetic or energetic barriers. In the original atomistic Trp-cage application, this clustering representation was based on time-lagged independent component analysis (TICA), which identifies linear combinations of molecular descriptors with slowly decaying time-lagged correlations \cite{Kania2026RiteWeight}. The same TICA representation used for the established MSM analysis was retained for RiteWeight to enable a controlled comparison between the two approaches. For the nonequilibrium folding analysis, this representation was constructed from minimal residue--residue distances using a 10-ns TICA lag time and 100 tICs. Although this representation facilitated comparison with the established MSM analysis, it was not specifically selected to optimize the RiteWeight clustering space.

This raises the question of whether the transition-relevant information
required by RiteWeight can be concentrated into a substantially
lower-dimensional representation. A compact feature space can reduce the cost
of repeated distance calculations and random clustering and, if kinetically
distinct configurations are more effectively separated, may also accelerate
RiteWeight convergence. Machine-learning-based dimensionality reduction
provides a natural route toward this goal because nonlinear representations can
be optimized directly using dynamical objectives rather than being restricted
to linear combinations of predefined molecular descriptors, as in conventional
TICA \cite{ZhangSchuette2023DeepLearningCV, Fu2024CollectiveVariableEnhancedSampling}. These considerations lead to two central questions. First, what type of
low-dimensional dynamical representation provides an effective clustering
space for RiteWeight? Second, can candidate representations be efficiently
selected using a common, representation-independent criterion without
performing a full downstream RiteWeight calculation for every
dimensionality-reduction method or hyperparameter set?

To address the first question, we investigate two complementary
representation-learning approaches with distinct dynamical objectives.
DeepTICA belongs to a class of operator-based methods that seek nonlinear
approximations to the leading nontrivial eigenfunctions of the molecular
transfer operator, which encode the dominant long-timescale dynamical
processes \cite{Bonati2021DeepTICA, PerezHernandez2013SlowMolecularOrderParameters}. Such coordinates may therefore provide an effective feature space
for separating kinetically distinct configurations during RiteWeight random
clustering. In contrast, the State Predictive Information Bottleneck (SPIB)
framework, referred to here as SPIB-VAE, learns a compressed latent
representation that retains information predictive of future metastable-state
identity \cite{WangTiwary2021SPIB}.

To address the second question, candidate representations were evaluated using
a coarse MSM-based kinetic score on held-out data, defined here as the sum of
the squared leading nontrivial MSM eigenvalues. The MSM was constructed using
a deliberately coarse discretization, with a clustering resolution comparable
to that used during RiteWeight random clustering. Such spectral kinetic scores
are commonly used in MSM model selection to identify representations and
discretizations that preserve the dominant slow dynamical processes and thereby
improve the quality of the finite-state Markov approximation at a chosen lag
time\cite{PerezHernandez2013SlowMolecularOrderParameters,McGibbonPande2015VariationalCV,
      HusicEtAl2016OptimizedMSM,
      WuNoe2020VAMP}. Here, however, the score is used for a different purpose. RiteWeight does
not require the coarse clusters themselves to define Markovian dynamics;
instead, the score serves as a proxy for whether kinetically distinct regions
remain sufficiently resolved in the reduced feature space used for random
clustering. Importantly, RiteWeight performance itself was not used during
model or hyperparameter selection. The subsequent RiteWeight calculations
therefore provide an independent test of whether this coarse-grained kinetic
criterion can efficiently identify suitable representation-learning methods
and their hyperparameters. Hyperparameters were explored using an initial
random search followed by Tree-structured Parzen Estimator (TPE) Bayesian
optimization.

Using atomistic Trp-cage dynamics as a test system, we use the
208~$\mu$s trajectory as equilibrium-like input and apply RiteWeight to
estimate the unfolded-to-folded mean first-passage time (MFPT) \cite{LindorffLarsen2011FastFolding}. RiteWeight
random clustering is performed separately in five-dimensional linear-TICA,
DeepTICA, and SPIB-VAE representations, allowing the effect of the reduced
feature space to be examined at fixed dimensionality. Candidate learned
representations are first evaluated using the held-out MSM-based kinetic
criterion and are subsequently tested as clustering spaces for RiteWeight over
a range of iteration counts. This framework allows us to assess both how the
dynamical information encoded by the reduced representation affects RiteWeight
convergence and whether a common coarse-grained kinetic criterion can be used
to efficiently select representation methods and their hyperparameters for
RiteWeight clustering.

\FloatBarrier

\section{Method}
\subsection{Trp-cage nonequilibrium data and state definitions}

We used the 208~$\mu$s atomistic MD trajectory of Trp-cage to study the
unfolded-to-folded mean first-passage time (MFPT), using the folded and unfolded
state definitions established in the original RiteWeight analysis~\cite{Kania2026RiteWeight}.
Briefly, minimal residue--residue distances were projected onto a TICA
representation constructed with a 10~ns lag time, and a 50-state reference MSM
was built in this space. PCCA++ analysis at a lag time of 100~ns was then used
to identify the two extreme states corresponding to folded and unfolded
conformations. These state definitions were retained throughout the present
study, with the unfolded state serving as the source and the folded state as
the sink in the RiteWeight calculations. The same folded and unfolded
macrostates were also used to construct the history-resolved haMSM reference
for evaluating the folding MFPT, as described in the Results section.

\subsection{DeepTICA representation}

We implemented a nonlinear reduced representation following the DeepTICA
framework\cite{Bonati2021DeepTICA}. The neural-network architecture, training
procedure, validation strategy, and hyperparameter-optimization workflow were
implemented in the present work. The \texttt{mlcolvar} package was used only
for the differentiable solution of the generalized TICA eigenvalue problem
required by the DeepTICA objective.

For each molecular configuration $X(t)$, a feed-forward neural network
$f_{\theta}$ generated a latent representation,
\begin{equation}
Z(t) = f_{\theta}[X(t)],
\end{equation}
where $\theta$ denotes the trainable neural-network parameters. The latent
representations at $t$ and $t+\Delta t$ were used to construct the
instantaneous covariance matrix $C_0$ and the symmetrized time-lagged
covariance matrix $C_{\Delta t}$. The generalized TICA eigenvalue problem,
\begin{equation}
C_{\Delta t} v_i = \lambda_i C_0 v_i,
\end{equation}
was solved using the differentiable \texttt{cholesky\_eigh} routine from the
\texttt{mlcolvar} package. Because this eigensolution remained within the
PyTorch computational graph, gradients of the eigenvalue-based objective could
be propagated through the neural network.

The network parameters were optimized by minimizing
\begin{equation}
\mathcal{L}_{\mathrm{DeepTICA}}
=
-\sum_{i=1}^{5}\lambda_i^2,
\end{equation}
so that training maximized the sum of the squared DeepTICA eigenvalues and
favored latent representations retaining slowly decorrelating dynamical
processes.

The input $X(t)$ consisted of minimal residue--residue distances. The neural
network contained two hidden layers with hyperbolic-tangent activation
functions followed by a linear output layer of dimension $d=5$. Because the
aim of this study was to compare different dimensionality-reduction approaches
as clustering representations for RiteWeight, the reduced-space dimensionality
was fixed at five for all methods. This ensured that differences in downstream
RiteWeight performance reflected differences in the organization of the
reduced feature space rather than differences in dimensionality.
The number of hidden layers and the activation function were fixed, whereas
the widths of the two hidden layers were optimized as described below.
Time-lagged training pairs were constructed using a lag time of 10~ns.

\subsection{DeepTICA hyperparameter optimization and model selection}

The DeepTICA training and model-selection workflow was implemented in PyTorch,
with hyperparameter optimization performed using Optuna\cite{optuna_2019}. Five quantities were
varied: the widths of the first and second hidden layers, $N_1$ and $N_2$; the
Adam learning rate, $\eta$; the minimum validation improvement required by the
early-stopping criterion, $\epsilon$; and the early-stopping patience, $p$. The corresponding search spaces were
\begin{equation}
N_1,N_2 \in \{16,32,64,128\},
\end{equation}
\begin{equation}
\eta \in [10^{-3},10^{-1}],
\end{equation}
with logarithmic sampling,
\begin{equation}
\epsilon \in [10^{-5},10^{-3}],
\end{equation}
also with logarithmic sampling, and
\begin{equation}
p \in \{20,40,80\}.
\end{equation}

Each candidate hyperparameter combination was evaluated independently over 10 train/validation splits. Within each split, the neural network was optimized on the training data using the DeepTICA eigenvalue objective. Validation-based early stopping was evaluated after an initial 150 training epochs. At each subsequent epoch, the DeepTICA eigenvalue objective was evaluated on the held-out data,
\begin{equation}
\mathcal{L}_{\mathrm{val}}
=
-\sum_{i=1}^{5}
\left(
\lambda_{i,\mathrm{val}}
\right)^2.
\end{equation}

A validation improvement was recorded when
\begin{equation}
\mathcal{L}_{\mathrm{val}}^{\mathrm{current}}
<
\mathcal{L}_{\mathrm{val}}^{\mathrm{best}}
-
\epsilon.
\end{equation}
When this condition was satisfied, the current network parameters and epoch were recorded and the patience counter was reset. Otherwise, the counter was incremented. Training was terminated when the counter reached $p$ or when the maximum of 1000 epochs was reached, after which the network parameters corresponding to the best validation epoch were restored. Thus, $\epsilon$ and $p$ were not parameters of the final DeepTICA representation itself; rather, they controlled the training duration and checkpoint selection for each candidate hyperparameter combination.

Importantly, the DeepTICA training objective itself was not used as the Optuna model-selection objective. Because the ultimate purpose of the learned representation was to provide a clustering space for RiteWeight, we instead evaluated the kinetic quality of each candidate representation using an independent MSM-based score on the held-out data. After restoring the best network checkpoint for each split, the TICA transformation was determined from the corresponding training data, and the held-out configurations were projected into the resulting five-dimensional DeepTICA space.

The held-out reduced features were partitioned into 10 states using $k$-means clustering. A MSM was then constructed from transitions between these states. Retention of slow kinetics was quantified using
\begin{equation}
S_{\mathrm{MSM}}
=
\sum_{i=1}^{7}
\left(
\lambda_i^{\mathrm{MSM}}
\right)^2,
\end{equation}
where the stationary eigenvalue was excluded and the first seven nontrivial MSM eigenvalues were included. In the implementation, Optuna minimized the equivalent negative score,
\begin{equation}
\mathcal{L}_{\mathrm{MSM}}
=
-
\sum_{i=1}^{7}
\left(
\lambda_i^{\mathrm{MSM}}
\right)^2.
\end{equation}
The score was averaged over the 10 train/validation splits, and the resulting mean value was used as the Optuna objective.

This model-selection criterion was motivated by the downstream use of the representation for RiteWeight clustering. RiteWeight repeatedly partitions the reduced feature space; therefore, a useful representation should distinguish configurations associated with slowly interconverting or barrier-separated regions. The MSM-based eigenvalue score provides a coarse-grained kinetic assessment of whether such slow processes remain resolved in the learned feature space. The subsequent RiteWeight calculations provide an independent test of whether hyperparameter selection based on this kinetic criterion yields a more effective clustering representation.

Hyperparameter sampling was performed using the Tree-structured Parzen Estimator (TPE) algorithm. The first 10 trials were sampled randomly, after which TPE was used to guide the remaining 40 trials, for a total of 50 trials per optimization run. A median-based Optuna pruner was additionally used to terminate poorly performing trials before completion of all 10 train/validation splits. After each completed split, the running mean MSM score was reported to Optuna, with pruning enabled only after the initial 10 startup trials.

\subsection{Final DeepTICA model}

After hyperparameter selection, a new DeepTICA model was trained using all available time-lagged data. The selected hidden-layer widths and learning rate were retained, whereas validation-based early stopping was no longer applied. Instead, the training duration was determined from the early-stopping results obtained during hyperparameter optimization. For the selected Optuna trial, the best validation epoch $e_{\mathrm{best}}^{(s)}$ was recorded independently for each of the 10 train/validation splits according to the early-stopping criterion described above. The final training duration was then chosen as
\begin{equation}
e_{\mathrm{run}}
=
\max
\left[
e_{\mathrm{best}}^{(1)},
e_{\mathrm{best}}^{(2)},
\ldots,
e_{\mathrm{best}}^{(10)}
\right].
\end{equation}
The final network was therefore trained on the complete time-lagged dataset for exactly $e_{\mathrm{run}}$ epochs.

After training, the neural network transformed the molecular descriptors into a five-dimensional nonlinear latent representation. TICA was then recomputed using the complete latent dataset, and the latent coordinates were projected onto the five resulting TICA eigenvectors. These five DeepTICA coordinates constituted the final reduced representation used for RiteWeight clustering.

\subsection{SPIB-VAE representation and hyperparameter optimization}

The information-bottleneck variational representation used here followed the
State Predictive Information Bottleneck (SPIB) framework\cite{WangTiwary2021SPIB}.
Briefly, SPIB encodes the current molecular configuration into a stochastic
latent representation and trains a decoder to predict the metastable-state
label at a future time, while a KL-divergence term regulates the information
retained in the latent space. We used the published nonlinear SPIB
implementation with iterative state-label refinement. As for DeepTICA, the
reduced-space dimensionality was fixed at five and time-lagged training pairs
were constructed using a lag time of 10~ns.

Hyperparameter optimization followed the same train/validation and MSM-based
model-selection procedure described above for DeepTICA. Seven SPIB
hyperparameters were varied: the encoder and decoder hidden-layer widths,
the learning rate, the information-bottleneck weight $\beta$, the
state-population convergence threshold, the convergence patience, and the
maximum number of label-refinement cycles. The corresponding search spaces
were
$N_1,N_2\in\{16,32,64,128\}$,
$\eta\in[10^{-5},5\times10^{-2}]$,
$\beta\in[10^{-5},10^{-1}]$,
the convergence threshold in $[10^{-3},10^{-1}]$,
$p\in\{1,2,3\}$, and
$n_{\mathrm{refine}}\in\{10,15,20\}$.
The learning rate, $\beta$, and convergence threshold were sampled
logarithmically.

Each candidate hyperparameter set was evaluated over the 10
train/validation splits similar as DeepTICA. The held-out five-dimensional
SPIB-VAE representation was clustered into 10 MSM states, and the same
MSM-based kinetic score constructed from the first seven nontrivial
eigenvalues was averaged across the splits and used as the Optuna objective.
Hyperparameter sampling likewise consisted of 10 initial random trials
followed by 40 TPE-guided trials, with the same median-based pruning strategy
described above.

\FloatBarrier

\section{Results}
\subsection{Effect of the reduced feature space on RiteWeight error}

RiteWeight performance was evaluated using the relative error in the folding mean first-passage time (MFPT) with respect to a history-augmented Markov state model (haMSM) reference. For both the RiteWeight and haMSM analyses, transition matrices were constructed from trajectory pairs separated by a lag time of 10~ns.

The haMSM model parameters used here were taken directly from those previously
established and validated for this Trp-cage dataset\cite{suarez2021markov}. In the haMSM
analysis, each trajectory segment is labeled according to whether the system
most recently visited the unfolded or folded macrostate. For estimation of the
folding MFPT, only trajectory segments whose most recently visited macrostate
was the unfolded state were used to construct the history-resolved source--sink
transition matrix, from which the corresponding nonequilibrium steady-state
distribution and folding MFPT were obtained as described previously\cite{suarez2021markov}.
Previous work has shown that this history-resolved construction provides
accurate estimates of source--sink steady-state populations and folding
kinetics\cite{suarez2021markov}.

RiteWeight is not provided these last-visited-state labels. Instead, it uses
the 10-ns trajectory pairs extracted from the equilibrium-like
208~$\mu$s Trp-cage trajectory and iteratively constructs weighted transition
matrices under unfolded-source/folded-sink boundary conditions. Random
clustering was applied only to configurations in the intermediate region,
excluding the predefined folded and unfolded states. The intermediate
configurations were partitioned into 10 random clusters, which together with
the folded and unfolded states yielded a $12\times12$ transition matrix.
To impose the source--sink condition, transitions from the folded sink were
redirected entirely to the unfolded source. The stationary solution of the
resulting transition matrix was then used to update the trajectory weights.

The folding mean first-passage time (MFPT) was then calculated from the
RiteWeight steady-state weights using the reciprocal-flux relation
\cite{Hill1989FreeEnergyTransduction,BhattZhangZuckerman2010SteadyStateWE},
\begin{equation}
\label{eq:mfpt}
\operatorname{MFPT}
=
\left(
\frac{1}{\tau}
\sum_{i_1 \notin \mathrm{folded},\, i_2 \in \mathrm{folded}}
w_i
\right)^{-1},
\end{equation}
where $\tau=10$~ns is the transition lag time and and $w_i$ is the weight assigned to each trajectory segment $i$. 

Across the calculations compared here, this RiteWeight procedure was kept
fixed; the only change was the reduced feature space in which the 10
intermediate random clusters were constructed. Specifically, clustering was
performed in five-dimensional linear-TICA, DeepTICA, or SPIB-VAE
representations.  The
resulting folding MFPT therefore provides a direct measure of how the choice of
reduced clustering representation affects RiteWeight convergence toward the
target source--sink steady state.

RiteWeight performance was quantified using the relative error with respect to
the haMSM reference MFPT,
\begin{equation}
\mathrm{Relative~Error}~(\%)
=
100\,
\frac{
\left|
\mathrm{MFPT}_{\mathrm{RW}}(n)
-
\mathrm{MFPT}_{\mathrm{haMSM}}
\right|
}{
\mathrm{MFPT}_{\mathrm{haMSM}}
},
\end{equation}
where $\mathrm{MFPT}_{\mathrm{RW}}(n)$ is the folding MFPT estimated after
$n$ RiteWeight iterations and $\mathrm{MFPT}_{\mathrm{haMSM}}$ is the
history-augmented reference value. Evaluating this error as a function of the
number of RiteWeight iterations allows the convergence efficiency of the
different reduced feature spaces to be compared directly. A lower relative error at a given iteration count indicates faster convergence of the folding MFPT toward the value associated with the target source--sink NESS.

\begin{figure}[t]
\centering
\includegraphics[width=1.0\linewidth, trim=0 0 0 0,clip]{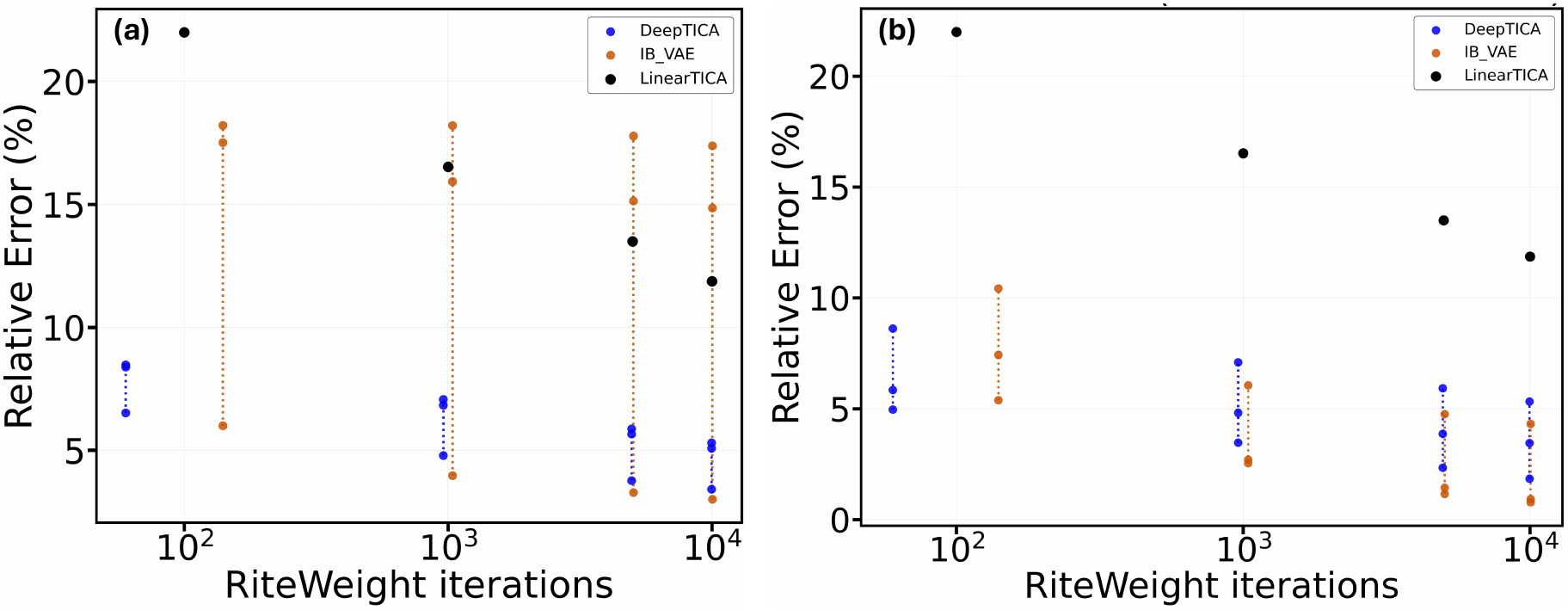}
\caption{Effect of the reduced feature representation on RiteWeight estimation of the
Trp-cage folding MFPT. Relative error with respect to the history-resolved
reference MFPT (HaMSM) is shown as a function of the number of RiteWeight iterations
for five-dimensional DeepTICA, SPIB-VAE, and linear-TICA clustering spaces.
For DeepTICA and SPIB-VAE, three independent hyperparameter-optimization runs
were performed. (a) Representations selected from the best of 10 randomly
sampled hyperparameter trials. (b) Representations selected from the best of
50 trials, consisting of 10 initial random trials followed by 40 TPE-guided
trials. Linear TICA provides a single baseline because no neural-network
hyperparameter optimization was performed.}
\label{fig:rw_feature_comparison}
\end{figure}

Figure~\ref{fig:rw_feature_comparison} compares the folding-MFPT relative
error obtained using five-dimensional linear TICA, DeepTICA, and SPIB-VAE
representations as the clustering space for RiteWeight. For the learned
representations, three independent hyperparameter-optimization runs were
performed, and each point corresponds to the representation selected from one
of these independent searches. The linear-TICA representation does not involve
neural-network hyperparameter optimization and therefore provides a single
reference value at each RiteWeight iteration count.

When the learned representations were selected using only 10 randomly sampled
hyperparameter combinations, DeepTICA consistently produced substantially
lower relative errors than the five-dimensional linear-TICA reference across
the tested RiteWeight iteration counts
(Fig.~\ref{fig:rw_feature_comparison}a). Moreover, the three independently
optimized DeepTICA representations yielded relatively similar errors,
indicating that an effective clustering representation could be obtained
without an extensive hyperparameter search. In contrast, the SPIB-VAE
representations exhibited considerably larger variability among the three
independent searches. Although some SPIB-VAE models achieved low error, other
hyperparameter selections produced substantially less accurate RiteWeight
estimates. Thus, under a limited search budget, DeepTICA was both more accurate
and more robust to hyperparameter selection.

Expanding the search to 50 trials, consisting of 10 initial random trials
followed by 40 TPE-guided trials, improved the performance of the learned
representations, with the largest improvement observed for SPIB-VAE
(Fig.~\ref{fig:rw_feature_comparison}b). Following the extended optimization,
both DeepTICA and SPIB-VAE produced substantially lower folding-MFPT errors than
five-dimensional linear TICA over the tested RiteWeight iteration counts.
The SPIB-VAE results also became markedly more consistent across independent
optimization runs and reached particularly low errors at the larger RiteWeight
iteration counts. DeepTICA improved more modestly, consistent with its already
strong performance after only the initial 10 random trials.

\subsection{Kinetic-score optimization and downstream RiteWeight performance}

\begin{figure}[t]
\centering
\includegraphics[width=1.0\linewidth, trim=0 0 0 0,clip]{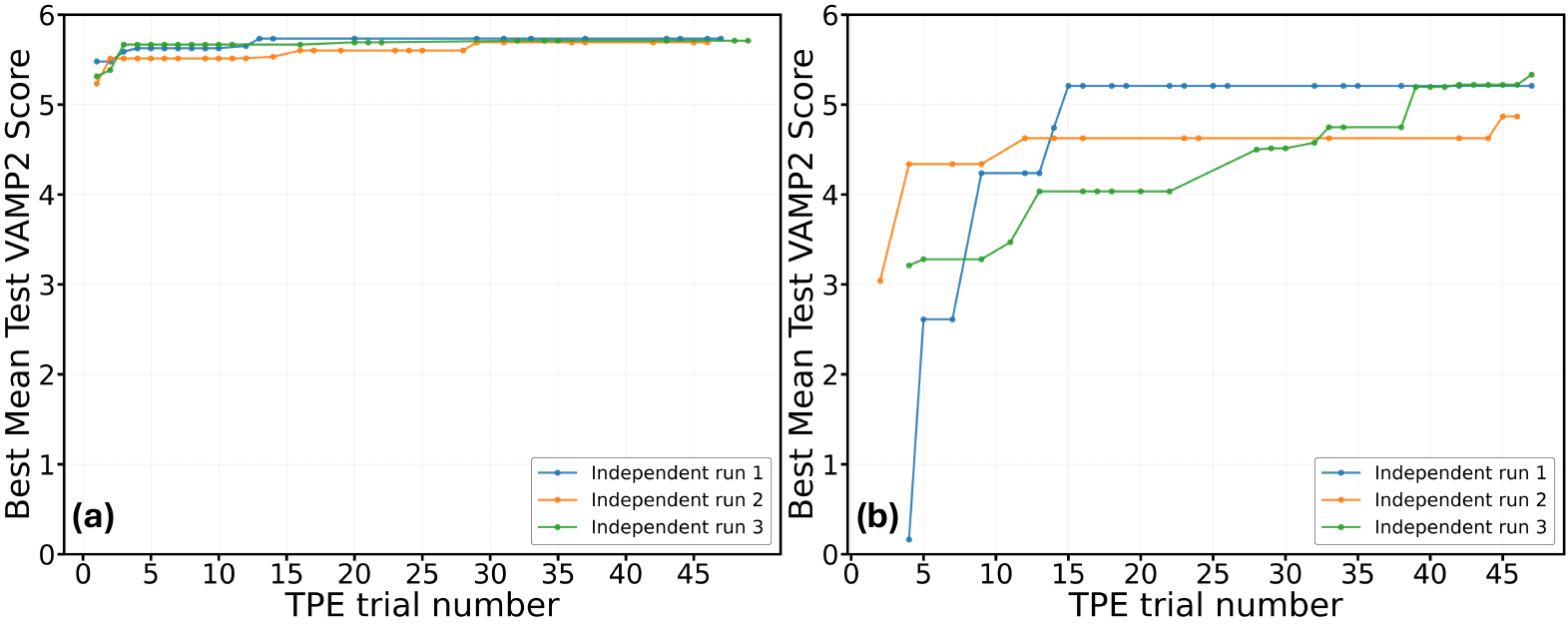}
\caption{
Progression of the best-so-far mean held-out MSM-based VAMP2 score during
hyperparameter optimization for (a) DeepTICA and (b) SPIB-VAE. For each
independent optimization run, the score shown at trial $n$ is the highest mean
held-out score obtained among all hyperparameter trials evaluated through trial
$n$. Scores were averaged over the 10 train/validation splits used to evaluate
each candidate hyperparameter set. The first 10 trials were sampled randomly,
followed by 40 trials selected using Tree-structured Parzen Estimator (TPE)
Bayesian optimization. DeepTICA reached a high-scoring region during the
initial random-search stage and showed only modest subsequent improvement,
whereas SPIB-VAE exhibited larger and more gradual gains during the extended
optimization.
}
\label{fig:kinetic_optimization}
\end{figure}

Figure~\ref{fig:kinetic_optimization} shows the progression of the
best-so-far mean held-out MSM kinetic score during hyperparameter optimization
for DeepTICA and SPIB-VAE. For each independent optimization run, the value shown
at a given trial number is the highest mean held-out score obtained among all
hyperparameter trials evaluated up to that point. The first 10 trials were
sampled randomly, after which the remaining 40 trials were selected using
Tree-structured Parzen Estimator (TPE) Bayesian optimization.

For DeepTICA, high kinetic scores were identified within the initial random
search, and only modest additional improvement was obtained during the
subsequent TPE-guided trials. In contrast, the SPIB-VAE searches showed larger
and more gradual increases in the best-so-far kinetic score as additional
hyperparameter combinations were evaluated. This behavior indicates that the
DeepTICA objective was relatively easy to optimize within the explored
hyperparameter space, whereas identification of high-scoring SPIB-VAE
representations benefited more strongly from the extended TPE search.

These optimization trends are qualitatively consistent with the downstream RiteWeight results in Fig.~\ref{fig:rw_feature_comparison}. DeepTICA already produced low RiteWeight errors when the representation was selected from only the first 10 random trials, consistent with the early saturation of its best-so-far kinetic score. In contrast, SPIB-VAE showed continued improvement in the kinetic score during the extended search and substantially better downstream RiteWeight performance after the full 50-trial optimization. Because RiteWeight error was not used during hyperparameter optimization, this agreement supports the held-out kinetic score from a coarse MSM with a clustering resolution comparable to RiteWeight as a practical criterion for selecting reduced representations and their hyperparameters.
\FloatBarrier
\section{Conclusion}

This study examined how the reduced feature space used for random clustering
affects the finite-iteration performance of RiteWeight. By fixing the reduced
dimensionality at five, we directly compared linear TICA with two nonlinear
dynamical representations, DeepTICA and SPIB-VAE, while keeping the remaining
RiteWeight procedure unchanged. For the Trp-cage folding problem, both learned
representations could provide substantially lower folding-MFPT errors than
five-dimensional linear TICA, demonstrating that the organization of the
reduced feature space can strongly influence RiteWeight convergence even when
the dimensionality itself is fixed.

The two learned approaches showed different sensitivities to hyperparameter
optimization. DeepTICA identified effective RiteWeight clustering spaces within
a relatively limited random search and improved only modestly during the
extended TPE-guided optimization. SPIB-VAE showed greater variability under the
limited search but benefited substantially from broader hyperparameter
exploration. These differences suggest that learning nonlinear representations
aligned with the leading nontrivial eigenfunctions of the molecular transfer
operator can provide a comparatively robust route for constructing compact
feature spaces for RiteWeight.

Importantly, representation selection was performed using only a held-out MSM-based kinetic score; RiteWeight error was not included in the hyperparameter-optimization objective. This score was obtained from a coarse MSM with a clustering resolution comparable to that used during RiteWeight random clustering, providing a coarse-grained assessment of whether kinetically distinct regions remain sufficiently resolved after dimensionality reduction. The qualitative correspondence between improvements in the held-out kinetic score and downstream RiteWeight performance therefore supports the use of such a coarse-grained kinetic criterion for selecting transition-relevant representations and their hyperparameters before performing the full iterative reweighting calculation.

Future work will be needed to determine how well these observations generalize across biomolecular systems and alternative molecular representations. In the present study, both DeepTICA and SPIB-VAE were constructed from predefined minimal residue--residue distance descriptors that were subsequently transformed by neural networks. An important extension would be to integrate graph-based architectures with DeepTICA to learn transition-relevant molecular representations more directly from three-dimensional structural information.

\section*{Acknowledgments}
The author acknowledges D. E. Shaw Research for making the
Trp-cage molecular dynamics trajectory used in this study available.

\bibliographystyle{unsrt}
\bibliography{references}

\end{document}